\documentclass[a4paper,10pt]{IEEEtran}
\renewcommand{\baselinestretch}{0.99}
\usepackage{cite}
\usepackage{amsmath,amsthm,amssymb,amsfonts,steinmetz}
\usepackage{graphicx,flushend}

\def\BibTeX{{\rm B\kern-.05em{\sc i\kern-.025em b}\kern-.08em
    T\kern-.1667em\lower.7ex\hbox{E}\kern-.125emX}}

\usepackage{tikz}
\usetikzlibrary{calc}
\makeatletter
\newcommand{\gettikzxy}[3]{%
  \tikz@scan@one@point\pgfutil@firstofone#1\relax
  \edef#2{\the\pgf@x}%
  \edef#3{\the\pgf@y}%
}
\makeatother

\usepackage{comment}
\usepackage{algorithm,algorithmic}
\usepackage{subcaption}

\renewcommand{\a}{\mathbf{a}}

\newcommand{\h}{\mathbf{h}}

\newcommand{\n}{\mathbf{n}}

\newcommand{\p}{\mathbf{p}}

\newcommand{\s}{\mathbf{s}}

\renewcommand{\u}{\mathbf{u}}
\renewcommand{\v}{\mathbf{v}}

\newcommand{\y}{\mathbf{y}}

\newcommand{\A}{\mathbf{A}}

\newcommand{\I}{\mathbf{I}}

\newcommand{\M}{\mathbf{M}}
\newcommand{\N}{\mathbf{N}}

\renewcommand{\P}{\mathbf{P}}

\newcommand{\W}{\mathbf{W}}

\newcommand{\Compl}{\mbox{$\mathbb{C}$}}

\renewcommand{\Im}{\mathrm{Im}}

\renewcommand{\Re}{\mathrm{Re}}

\DeclareMathAlphabet\mathbfcal{OMS}{cmsy}{b}{n}

\begin{document}
\title{Circuit-Compliant Optimization of Dynamic Metasurface Antennas for Near-Field Localization}
\author{\IEEEauthorblockN{
Ioannis Gavras$^1$ and George C. Alexandropoulos$^{1,2}$ 
} \\
\IEEEauthorblockA{$^1$Department of Informatics and Telecommunications, National and Kapodistrian University of Athens, Greece}\\
\IEEEauthorblockA{$^2$Department of Electrical and Computer Engineering, University of Illinois Chicago, USA}\\
\IEEEauthorblockA{emails: \{giannisgav, alexandg\}@di.uoa.gr}

}
\pagenumbering{gobble}
\maketitle

\begin{abstract}
This paper presents an optimization framework for near-field localization with Dynamic Metasurface Antenna (DMA) receivers. This metasurface technology offers enhanced angular and range resolution realizing efficient hybrid Analog and Digital (A/D) BeamForming (BF) with sub-wavelength-spaced metamaterials of tunable responses. However, the vast majority of the state-of-the-art DMA designs is based on an idealized model for their reception operation, which neglects several practical aspects, such as the inevitable mutual coupling among the densely deployed metamaterials within a given aperture. Capitalizing on a recent circuit-compliant active metasurface model, we present a novel mutual-coupling-aware framework for localization-optimized hybrid A/D BF weights at the reception DMA. To deal with the intrinsic complexity of the deployed model, we introduce first- and second-order approximations for the DMA analog BF matrix that enable efficient optimization, while maintaining accuracy. We derive the Cramér-Rao Bound for the user position estimation which serves as our design objective for the hybrid A/D BF matrices. Closed-form solutions for these matrices for both approximations are presented, whose validity is confirmed via numerical investigations. It is also demonstrated that the proposed DMA design outperforms state-of-the-art multi-antenna reception architectures optimized for the same localization objective. 
\end{abstract}


\section{Introduction}
Extremely large (XL)  antenna arrays are expected to be widely deployed in next generation wireless networks~\cite{XLMIMO_tutorial,41}, offering extensive spatial degrees of freedom for communications, localization, and sensing applications. Dynamic Metasurface Antennas (DMAs), a special form of the technology of reconfigurable intelligent surfaces~\cite{BAL2024,ASA2024}, constitute a recent power- and cost-efficient hybrid Analog and Digital (A/D) transceiver architecture utilizing arbitrary large numbers of phase-tunable metamaterials, which are grouped in disjoint microstrips, each attached to a Radio Frequency (RF) chain and are capable for realizing digital and analog transmit or Receive (RX) BeamForming (BF)~\cite{Shlezinger2021Dynamic,YXA2023,gavriilidis2024metasurface,NF_beam_tracking}.  

Metasurface-based antennas have recently garnered significant research interest, leading to numerous studies exploring their diverse applications and use cases. The beam focusing capability of a DMA-based transmitter serving multiple User Equipments (UEs) within its near-field regime was optimized in~\cite{zhang2022beam}. A localization framework for DMA-based RXs with $1$-bit quantized hybrid A/D reception was developed in~\cite{gavras2024near}. In~\cite{Nlos_DMA}, an autoregressive attention neural network was introduced for non-line-of-sight user tracking using a DMA-based RX, while a near-field beam-tracking framework for the same RX architecture was presented in~\cite{NF_beam_tracking}. Additionally, a method for configuring the analog DMA weights to improve near-field localization was proposed in~\cite{yang2023near}, transforming the localization objective into a signal strength maximization problem. More recently,~\cite{FD_HMIMO_2023,spawc2024,gavras2024joint} presented efficient designs for the A/D BF matrices of DMA-based full duplex XL transceivers in near-field THz scenarios, aiming simultaneous sensing and communication~\cite{IAS2022b}. However, all of the aforementioned works and the vast majority of open technical literature rely on idealized DMA models that disregard practical effects, such as mutual coupling, which is known to be present in sub-wavelength spaced antenna arrays~\cite{akrout2023physically}. To the best of the authors' knowledge, no studies have yet utilized an electromagnetic-compliant DMA model that accounts for mutual coupling among the active elements of the metasurface to design efficient communication and localization schemes.

Motivated by this research gap, this paper capitalizes on the recent circuit-compliant DMA model described in~\cite{williams2022electromagnetic} and presents a novel mutual-coupling-aware optimization framework for single-user localization with metasurface-based reception. Due to the complexity of the considered model, a convenient simplification, whose accuracy is verified via extensive numerical investigation, is introduced that facilitates optimization of the RX parameters. Based on this approximate DMA model, we derive the Cramér-Rao Bound (CRB) for the estimation of the user range and angular parameters, as well as the Position Error Bound (PEB). We then consider the minimization of the CRB as the objective for designing the A/D BF weights of the DMA-based RX. Leveraging the structure of the DMA, we reformulate the design objective as a Rayleigh quotient optimization problem and present closed-form A/D BF solutions for the first- and second-order approximations of the proposed DMA model. Our simulation results validate the accuracy of our localization analysis, showcase the effects of mutual coupling on the localization performance, and demonstrate the benefits of our proposed CRB-minimizing DMA design over established state-of-the-art multi-antenna RX architectures that have been designed to overlook mutual coupling between adjacent antenna/metamaterial elements.

\textit{Notations:}
Vectors and matrices are represented by boldface lowercase and uppercase letters, respectively. The transpose, Hermitian transpose, inverse, and Euclidean norm are denoted as $(\cdot)^{\rm T}$, $(\cdot)^{\rm H}$, $(\cdot)^{-1}$, and $\|\cdot\|$, respectively. $[\mathbf{A}]_{i,j}$ and $[\mathbf{A}]_{i:j,u:v}$ give respectively $\mathbf{A}$'s $(i,j)$th element and sub-matrix spanning rows $i$ to $j$ and columns $u$ to $v$. $\mathbf{I}_{n}$ and $\mathbf{0}_{n}$ ($n\geq2$) are the $n\times n$ identity and zeros' matrices, respectively, and $\boldsymbol{1}_{N}$ is an $N \times 1$ column vector of ones. 
$\mathbb{C}$ is the complex number set, $|a|$ is the amplitude of scalar $a$, and $\jmath\triangleq\sqrt{-1}$. $\mathbb{E}\{\cdot\}$, ${\rm Tr}\{\cdot\}$, and $\Re\{\cdot\}$ give the expectation, trace, and real part, respectively. 
Finally, $\mathbf{x}\sim\mathcal{CN}(\mathbf{a},\mathbf{A})$ indicates a complex Gaussian random vector with mean $\mathbf{a}$ and covariance matrix $\mathbf{A}$.

\section{System and Channel Models}
We consider an RX equipped with an XL DMA panel~\cite{Shlezinger2021Dynamic}, which is tasked to localize a single-antenna User Equipment (UE) that, for this purpose, transmits pilot signals in the uplink (UL) direction. DMAs efficiently enable the integration of numerous sub-wavelength-spaced metamaterials of tunable responses, which are usually grouped in microstrips, each attached to a reception RF chain (comprising a low noise amplifier, a mixer downconverting the received signal from RF to baseband, and an analog-to-digital converter), within possibly XL apertures~\cite{41}. We assume that the RX DMA panel is situated in the $xz$-plane with the first microstrip positioned at the origin. There exist in total $N_{\rm RF}$ microstrips within the panel, each composed of $N_{\rm E}$ distinct metamaterials with $d_{\rm E}$ distance between adjacent elements. The microstrips are individually linked to reception RF chains, which are separated from one another by a distance of $d_{\rm RF}$. Consequently, the RX DMA includes in total $N\triangleq N_{\rm RF}N_{\rm E}$ metamaterials.

The received symbols at the RX undergo initial analog processing through the analog BF matrix $\W_{\rm RX}\in \Compl^{N_{\rm RF}\times N}$ with the tunable absorption states of all metamaterials, followed by digital processing via the vector $\v \in\Compl^{N_{\rm RF} \times 1}$. Following the circuital DMA model of~\cite{williams2022electromagnetic}, $\W_{\rm RX}$ captures the relationship between the received pilot symbols at the output of the reception RF chains and their analog processing taking place inside each microstrip, accounting for the wave propagation and reflections throughout the waveguides that feed the RF chains, and the mutual coupling both through the air and those waveguides. To this end, the analog BF matrix $\W_{\rm RX}$ can be expressed as follows:
\begin{align}\label{eq: BF}
    \W_{\rm RX} = \P_{\rm SA}^{\rm H}\left(\W_{\rm TA}+\W_{\rm MC}\right)^{-1},
\end{align}
where $\P_{\rm SA} \in \Compl^{N \times N_{\rm RF}}$ represents signal propagation from each reception RF chain to its associated metamaterial elements within their respective waveguides. Consequently, its entries are zero for element pairs in different waveguides, as they lack a shared RF chain. In particular, the entries of $\P_{\rm SA}$ can be calculated $\forall$$n=1,2,\ldots,N$ and $\forall$$i=1,\ldots,N_{\rm RF}$ as:
\begin{align}
    \nonumber&[\P_{\rm SA}]_{n,i} \\ \nonumber&=\begin{cases}
    \jmath\omega{G}_{\rm SA}(\p_n,\p_i),&  \text{for $\p_n,\p_i$ in the same waveguides}\\
    0,              & \text{for $\p_n,\p_i$ in different waveguides}
\end{cases},
\end{align}
where $\p_n\triangleq [x_n, y_n, z_n]$ and $\p_i\triangleq [x_i, y_i, z_i]$ denote the three Dimensional (3D) cartesian coordinates of each $n$-th element connected to each $i$-th RX RF chain and the $i$th RX RF chain output port. Additionally, ${G}_{\rm SA}(\p_n,\p_i)$ represents the third diagonal component of the Green’s function inside the waveguide~\cite{williams2022electromagnetic} and captures the waveguide propagation between any pair of points $\p_n$ and $\p_i$, and is given as:
\begin{align}
    \nonumber G_{\rm SA}&\left(\p_n, \p_i\right)=\frac{-k_x \sin \left(\frac{\pi z_n}{a}\right) \sin \left(\frac{\pi z_i}{a}\right)}{a b k^2 \sin \left(k_x S_\mu\right)} \\\nonumber&\times\left[\cos \left(k_x\left(x_i+x_n-S_\mu\right)\right)+\cos \left(k_x\left(S_\mu-\left|x_n-x_i\right|\right)\right)\right],
\end{align}
where $a, b$, and $S_{\mu}$ denote the waveguide’s width, height, and length, respectively. Furthermore, $k_x$ is defined as $k_x\triangleq\Re\{\sqrt{k^2-(\pi/a)^2}\}-\jmath\Im\{\sqrt{k^2-(\pi/a)^2}\}$, where $k$ is the waveguide’s wavenumber that is given as $k\triangleq2\pi/(\lambda\sqrt{\epsilon_r\mu_r})$ with $\lambda$,
$\epsilon_r$, and $\mu_r$ being the wavelength of operation, the relative permittivity, and the relative permeability, respectively.

A key challenge in the ultra-dense DMA architectures is the mutual coupling between adjacent radiating elements due to their sub-wavelength spacing. This coupling occurs both through the air between surface elements and within the waveguide for elements in the same waveguide. To this end, each $(n,n')$-th entry of $\W_{\rm MC} \in \Compl^{N \times N}$ $\forall$$n,n'=1,2,\ldots,N$, can be calculated for $(\p_n,\p_{n'})$ pair in the same waveguide (with $\p_n$ and $\p_{n'}$ being the 3D cartesian coordinates for each $n$-th and $n'$-th metamaterial, respectively) as follows~\cite{williams2022electromagnetic}:
\begin{align}
    \nonumber(\jmath\omega\epsilon)^{-1}[\W_{\rm MC}]_{n,n'} = 2G_{\rm MC}(\p_n,\p_{n'})+G_{\rm SA}(\p_n,\p_{n'}),
\end{align}
whereas, for $\p_n$ and $\p_{n'}$ placed in different waveguides, $(\jmath\omega\epsilon)^{-1}[\W_{\rm MC}]_{n,n'}=2G_{\rm MC}(\p_n,\p_{n'})$. In the latter expression, $\epsilon$ is the medium’s permittivity and $G_{\rm MC}(\p_n,\p_{n'})$ is the third diagonal component of the Green’s function in free space, which is given by:
\begin{align}
    \nonumber &G_{\rm MC}(\p_n,\p_{n'})  \\\nonumber&=\left(\frac{R^2-\Delta z^2}{R^2}-\jmath\frac{R^2-3\Delta z^2}{R^3k}\right)\left(\frac{R^2-3\Delta z^2}{R^4k^2}\right)\frac{e^{-\jmath kR}}{4\pi R},
\end{align}
where $R \triangleq \|\p_n-\p_{n'}\|$ and $\Delta z\triangleq z_n-z_{n'}$.

Finally, in~\eqref{eq: BF}, $\W_{\rm TA}\in\Compl^{N\times N}$ is a diagonal matrix whose each non-zero element $[\W_{\rm TA}]_{n,n}$ denotes the termination admittance of each $n$-th metamaterial. The inverse of the termination admittances are considered to follow the Lorentzian constrained profile as in \cite{zhang2022beam,FD_HMIMO_2023}, herein, we assume that they belong to a phase profile codebook $\mathcal{W}$, as follows:
\begin{align}\label{eq: Code}
    \left[\W_{\rm TA}^{-1}\right]_{n,n}\in\mathcal{W}\triangleq\left\{0.5(\jmath+e^{\jmath\phi_n})\Big|\phi_n\in\left[-\frac{\pi}{2},\frac{\pi}{2}\right]\right\}.
\end{align}

\subsection{DMA Model Approximation}\label{Sec: MC_approx}
In the widely adopted idealized DMA model for the analog BF matrix, $\W_{\rm RX}$ becomes a block diagonal matrix including instances of the $\W_{\rm TA}$ matrix, thus, it can be easily optimized for various objectives~\cite{Shlezinger2021Dynamic,YXA2023,gavriilidis2024metasurface,NF_beam_tracking,zhang2022beam,gavras2024near,Nlos_DMA,yang2023near,FD_HMIMO_2023,spawc2024,gavras2024joint,IAS2022b}. However, the circuital model expression for $\W_{\rm RX}$ in~\eqref{eq: BF} is quite challenging due to the inverse of the sum of the termination
admittance and mutual coupling matrices. To tackle this difficulty, we approximate~\eqref{eq: BF}, similar to~\cite{FSA2023,RLS2024}, to enable a more tractable system design and optimization. To this end, utilizing the Woodbury matrix identity formula with respect to $\W_{\rm MC}$, we can expand the inverse of the matrix included in~\eqref{eq: BF} as follows:
\begin{align}
    \nonumber\left(\W_{\rm TA}+\W_{\rm MC}\right)^{\rm -1} &= \W_{\rm MC}^{\rm -1}-\W_{\rm MC}^{\rm -1}\left(\W_{\rm MC}\W_{\rm TA}^{\rm -1}+\I_N\right)^{\rm -1}.
\end{align}
Next, we apply the Taylor series expansion to the term $\left(\W_{\rm MC}\W_{\rm TA}^{\rm -1}+\I_N\right)^{\rm -1}$, yielding:
\begin{align}
    \nonumber\left(\W_{\rm MC}\W_{\rm TA}^{\rm -1}+\I_N\right)^{\rm -1} \approx \I_N+\sum_{n=1}^{\infty}(-1)^n\left(\W_{\rm MC}\W_{\rm TA}^{\rm -1}\right)^n.
\end{align}
Combining the previously derived expressions results in:
\begin{align}
    \nonumber\left(\W_{\rm TA}+\W_{\rm MC}\right)^{\rm -1}\approx -\W_{\rm MC}^{-1}\sum_{n=1}^{\infty}(-1)^n\left(\W_{\rm MC}\W_{\rm TA}^{\rm -1}\right)^n.
\end{align}
\begin{figure}[!t]
	\begin{center}
	\includegraphics[scale=0.55]{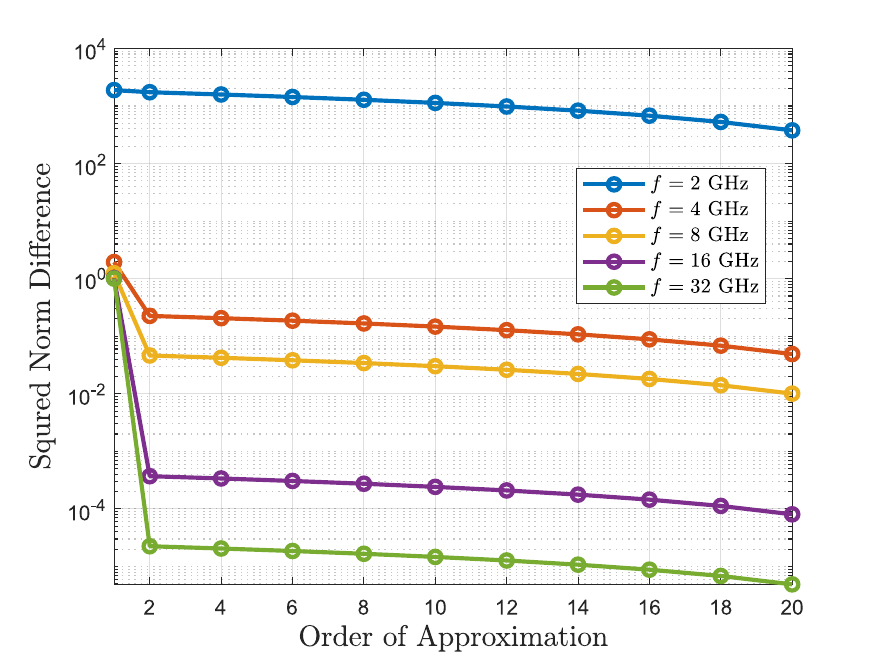}
	\caption{\small{Squared norm difference between the approximated and actual values of $(\W_{\rm TA}+\W_{\rm MC})^{-1}$ as a function of the approximation order for various operating frequencies $f$, considering an RX DMA panel with $N_{\rm RF}=4$ microstrips each with $N_{\rm E}=32$ metamaterials.}}
    \vspace{-0.4cm}
	\label{fig: approx}
	\end{center}
\end{figure}

In Fig.~\ref{fig: approx}, we illustrate the approximation error in terms of the squared norm difference between the approximated and actual values of $\left(\W_{\rm TA}+\W_{\rm MC}\right)^{-1}$ versus the approximation order for various frequencies of operation, considering an RX DMA panel with $N_{\rm RF}=4$ RX RF chains, each composed by $N_{\rm E}=32$ response-tunable metamaterials. We have assumed that the spacing between different RF chains and between adjacent metamaterials within a waveguide is $d_{\rm RF}=\lambda$ and $d_{\rm RF}=\lambda/5$, respectively, where $\lambda=c/f$ denotes the wavelength with $c$ representing the speed of light and $f$ the operational frequency. The rest of the RX DMA's modeling parameters were taken from \cite[Table II]{williams2022electromagnetic}. It is evident from the figure that, as the value of $f$ increases, each additional successive approximation order larger than $2$ slightly improves the approximations accuracy. This implies that, retaining only the first few terms, results in accurate approximations, though, additional terms add complexity with minimal benefit. Therefore, utilizing just up to the first or the second term in \eqref{eq: BF}'s approximation can be deemed sufficient. We have also noticed that the same trend holds when increasing the number of metamaterials per RX RF chain or the number of RX RF chains at a fixed frequency and spacing. 

Interestingly, when considering only the first-order approximation, yields $(\W_{\rm TA}+\W_{\rm MC})^{\rm -1} \approx \W_{\rm TA}^{\rm -1}$ and, thus, $\W_{\rm RX} \approx \P_{\rm SA}^{\rm H}\W_{\rm TA}^{-1}$, which significantly simplifies the analog BF expression, while entirely neglecting mutual coupling effects. In the following section, we address the problem of PEB minimization and present BF design solutions based on the first- and second-order approximations of~\eqref{eq: BF}. These approaches correspond to optimization scenarios where mutual coupling effects are either ignored or incorporated.

\subsection{Channel Model}
We consider a UE located in the vicinity of the DMA panel, characterized by the polar coordinates $(r,\theta,\varphi)$, representing respectively the UE's distance from the origin, its elevation and azimuth angles. 
The $N$-element complex-valued vector channel between the RX DMA and the single-antenna UE is modeled in the near-field region as follows:
\begin{align}
    \label{eqn:UL_chan}
    [\h]_{(i-1)N_{\rm E}+n} \triangleq \sqrt{F(\theta_{i,n})} \frac{\lambda}{4\pi r_{i,n}}  \exp\left(\frac{\jmath2\pi}{\lambda} r_{i,n}\right),
\end{align}
where $F(\cdot)$ denotes each metamaterial's radiation profile \cite[eq. (6)]{FD_HMIMO_2023}, $r_{i,n}$ denotes the distance from the UE's antenna to each $n$-th metamaterial of each $i$-th microstrip (and respective reception RF chain) of the RX DMA panel, and $\theta_{i,n}$ is the corresponding elevation angle. It is finally noted that the distance $r_{i,n}$ in~\eqref{eqn:UL_chan} can be calculated as:
\begin{align}\label{eq: dist}
    \nonumber r_{i,n} =& \Big(\!(r\sin\theta\cos\phi -(i\!-\!1)d_{\rm RF})^2 +\\ &(r\sin\theta\sin\phi)^2 + (r\cos\theta\!-\!(n\!-\!1)d_{\rm E})^2\Big)^{\frac{1}{2}}.
\end{align}
The elevation angle of the UE's antenna with respect to each $n$-th element of each $i$-th microstrip is expressed as follows:
\begin{align}\label{eq:thetas}
    \theta_{i,n} \triangleq \sin^{-1}\left(r_{i,n}^{-1}|(n-1)d_{\rm E}-r\cos{\theta}|\right).
\end{align}

\subsection{Received Signal Model}
The digitally processed baseband received signal at the output of the DMA's reception RF chains after $T$ UE pilot transmissions can be mathematically expressed via the matrix $\y\triangleq [y(1),y(2),\ldots,y(T)]\in\Compl^{1\times T}$ with $t=1,2,\ldots,T$:
\begin{align}\label{eq:UL_signal_matrix}
    \y \triangleq \v^{\rm H}\W_{\rm RX}\h^{\rm H}\s+\v^{\rm H}\W_{\rm RX}\N,
\end{align}
where $\N \triangleq [\n(1),\n(2),\ldots,\n(T)]\in\Compl^{N\times T}$ with each $\n(t)\sim\mathcal{CN}(\mathbf{0},\sigma^2\mathbf{I}_{N})$ being the Additive White Gaussian Noise (AWGN) vector, and $\s \triangleq [s(1),s(2),\ldots,s(T)]\in\Compl^{1\times T}$. Each transmitted pilot signal $s(t)$ is subjected to the power constraint $\mathbb{E}\{\|s(t)\|^2\}\leq P_{\max}$, where $P_{\max}$ signifies the maximum UE transmission power in the UL direction. 

\section{DMA Optimization for Localization}
In this section, we present our RX DMA design for near-field localization. We first derive the CRB and PEB for the intended UE estimation, which are then used for our design objective formulation. Then, our closed-form solutions for the DMA's A/D BF weights are presented.

\subsection{PEB Analysis}
It is evident from  \eqref{eq:UL_signal_matrix}'s inspection that, for a sufficiently large number of UE pilot transmissions $T$ within a coherent channel block, yields $T^{-1}\mathbb{E}\{\s\s^{\rm H}\}= P_{\max}$, indicating that the digitally processed signal at the outputs of the RX RF chains of the DMA is distributed as $\y\sim\mathcal{CN}(\boldsymbol{\mu},\boldsymbol{\Sigma})$ with mean $\boldsymbol{\mu} \triangleq \v^{\rm H}\W_{\rm RX}\h^{\rm H}\s$ and covariance $\boldsymbol{\Sigma} \triangleq\sigma^2\v^{\rm H}\W_{\rm RX}\W_{\rm RX}^{\rm H}\v\I_T$. Considering the estimation of the UE's position parameter vector $\boldsymbol{\zeta} \triangleq [r,\theta,\varphi]^{\rm T}$, the $3\times3$ Fisher Information Matrix (FIM) can be calculated as follows~\cite{kay1993fundamentals}: 
\begin{align}\label{eq:FIM}
    \mathbfcal{I} \triangleq\begin{bmatrix}
    \mathbfcal{I}_{rr} & \mathbfcal{I}_{r\theta} & \mathbfcal{I}_{r\phi}\\
    \mathbfcal{I}_{\theta r} & \mathbfcal{I}_{\theta\theta} & \mathbfcal{I}_{\theta\phi}\\
    \mathbfcal{I}_{\phi r} & \mathbfcal{I}_{\phi\theta} & \mathbfcal{I}_{\phi\phi}
    \end{bmatrix},
\end{align}
where each $(i,j)$-th element of $\boldsymbol{\mathcal{I}}$ is formulated as:
\begin{align}
    \nonumber[\mathbfcal{I}]_{i,i}\!=\!2\Re\left\{\frac{\partial \boldsymbol{\M}^{\rm H}}{\partial[\boldsymbol{\zeta}]_i}\boldsymbol{\Sigma}^{-1}\frac{\partial \boldsymbol{\M}}{\partial[\boldsymbol{\zeta}]_j}\right\}\!+\!\text{Tr}\left\{\boldsymbol{\Sigma}^{-1}\frac{\partial\boldsymbol{\Sigma}}{\partial[\boldsymbol{\zeta}]_i}\boldsymbol{\Sigma}^{-1}\frac{\partial\boldsymbol{\Sigma}}{\partial[\boldsymbol{\zeta}]_j}\right\}.
\end{align}
Since $\nabla_{\boldsymbol{\zeta}}\boldsymbol{\Sigma} = \mathbf{0}_{3\times1}$ because $\boldsymbol{\Sigma}$ is independent of $\boldsymbol{\zeta}$, it holds that each FIM value depends solely on $\y$'s mean value. We next compute the following derivative $\forall i$:
\begin{align}\label{eq: deriv_mean}
    \frac{\partial\boldsymbol{\mu}}{\partial[\boldsymbol{\zeta}]_i} = \v^{\rm H}\W_{\rm RX}\frac{\partial\h^{\rm H}}{\partial[\boldsymbol{\zeta}]_i}\s.
\end{align}
Consequently, each diagonal element of the FIM matrix in \eqref{eq:FIM} can be expressed as follows:
\begin{align}\label{eq: FIM_diag}
    [\mathbfcal{I}]_{i,i}=\frac{2}{\sigma^2}\Re\left\{\frac{\s^{\rm H}\frac{\partial\h}{\partial[\boldsymbol{\zeta}]_i}\W_{\rm RX}^{\rm H}\v\v^{\rm H}\W_{\rm RX}\frac{\partial\h^{\rm H}}{\partial[\boldsymbol{\zeta}]_i}\s}{\v^{\rm H}\W_{\rm RX}\W_{\rm RX}^{\rm H}\v}\right\}.
\end{align}
Putting all above together, the PEB for a single-antenna UE with polar coordinates $(r,\theta,\phi)$ and an RX DMA is given by:
\begin{align}\label{eq:PEB}
\text{PEB}_{\boldsymbol{\zeta}} \triangleq \sqrt{{\rm CRB}_{\boldsymbol{\zeta}}}=
\sqrt{{\rm Tr}\left\{\mathbfcal{I}^{-1}\right\}}. 
\end{align}

\subsection{DMA Design Objective}
Our goal, in this paper, is to optimize the RX DMA's A/D BF weights for UE localization. To this end, we leverage the positive semidefinite nature of the FIM and the lower bound ${\rm Tr}\{\mathbfcal{I}^{-1}\}\geq\frac{9T^2}{{\rm Tr}\{\mathbfcal{I}\}}$ that results from the previously derived PEB expression in~\eqref{eq:PEB}.  We then focus on the minimization of a tight lower bound for the estimation of the UE position parameter  $\boldsymbol{\zeta}$, which can be formulated via the optimization:
\begin{align}
        \mathcal{OP}&:\nonumber\underset{\substack{\W_{\rm RX},\v}}{\max} \, \frac{\v^{\rm H}\W_{\rm RX}\A\W_{\rm RX}^{\rm H}\v}{\v^{\rm H}\W_{\rm RX}\W_{\rm RX}^{\rm H}\v}\,\,\text{\text{s}.\text{t}.}\,
        \left[\W_{\rm TA}^{-1}\right]_{n,n}\in\mathcal{W}\,\,\forall n,
\end{align}
where $\A \triangleq \sum_{i=1}^3\frac{\partial\h^{\rm H}} {\partial[\boldsymbol{\zeta}]_i}\frac{\partial\h}{\partial[\boldsymbol{\zeta}]_i}$ with $\A\succeq0$ (i.e, a positive semidefinite matrix). The constraint ensures that the analog BF weights satisfy the Lorentzian form constraint in \eqref{eq: BF}. By defining $\widetilde{\v}\triangleq\W_{\rm RX}^{\rm H}\v$, the unconstrained version of $\mathcal{OP}$ can be expressed as follows:
\begin{align}\
        &\nonumber\underset{\substack{\widetilde{\v}}}{\max} \quad \frac{\widetilde{\v}^{\rm H}\A\widetilde{\v}}{\|\widetilde{\v}\|^2}.
\end{align}
In particular, omitting the Lorentzian form constraint which is non-convex, transforms the original problem $\mathcal{OP}$ into a Rayleigh quotient maximization problem. This formulation has a well-known rank-one optimal solution $\widetilde{\v}_{\rm opt}\triangleq \u_1\sqrt{\sigma_1}$, where $\sigma_1$ is the largest singular value of $\A$ and $\u_1$ denotes its corresponding singular vector.

We now need to decompose the optimal rank-one solution into the desired A/D BF solutions. This can be achieved in multiple ways, but because of the Lorentzian constraint imposed by the $\W_{\rm TA}^{-1}$'s elements, we propose to minimize the difference between the optimal solution and the product of the A/D BF matrices in the least squares sense \cite{zhang2022beam}. The former can be mathematically formulated as follows:
\begin{align}
        \mathcal{OP}_1:&\nonumber\underset{\substack{\W_{\rm TA},\v}}{\min} \left\|\widetilde{\v}_{\rm opt}-\W_{\rm RX}^{\rm H}\v\right\|^2\,\,\text{\text{s}.\text{t}.}\, \left[\W_{\rm TA}^{-1}\right]_{n,n}\in\mathcal{W}\,\,\forall n.
\end{align}
For a given matrix $\W_{\rm TA}$, the least-squares solution $\v=(\W_{\rm RX}\W_{\rm RX}^{\rm H})^{-1}\W_{\rm RX}\widetilde{\v}_{\rm opt}$ is known to optimally minimize the norm difference. In the subsequent analysis, we derive the closed-form solutions for the analog BF weight matrix $\W_{\rm TA}^{-1}$, based on the first- and second-order approximations of \eqref{eq: BF}, as previously presented in Section~\ref{Sec: MC_approx}.

\subsubsection{First-Order Approximation}\label{eq: ex1}
Considering the previously derived first-order approximation, we have that $\W_{\rm RX}\approx\P_{\rm SA}^{\rm H}\W_{\rm TA}^{-1}$. By defining $\a \triangleq \P_{\rm SA}\v$, we reformulate $\mathcal{OP}_1$'s objective with respect to each $n$-th diagonal value $w_n \triangleq \left[\W_{\rm TA}^{-1}\right]_{n,n}=0.5(\jmath+e^{\jmath\phi_n})$ of the termination
admittance matrix as follows:
\begin{align}\label{eq: f_W}
    \nonumber&f_1(w_n) \triangleq \left|[\widetilde{\v}_{\rm opt}]_n-\left[\W_{\rm TA}^{-1}\right]_{n,n}^{\rm H}[\a]_n\right|^2
     \\& = \left|\left[\widetilde{\v}_{\rm opt}\right]_n\right|^2-2\Re\{w_n^*\left[\widetilde{\v}_{\rm opt}^*\right]_n[\a]_n\}+|w_n|^2|[\a]_n|^2.
\end{align}
Next, we further expand \eqref{eq: f_W} as in \eqref{eq: tt2} (top of the next page) by incorporating the Lorentzian form of the analog BF weight $w_n$. We then compute the derivative of $f_1(\phi_n)$ with respect to the phase shift $\phi_n$, as shown in \eqref{eq: tt3}. Finally, equating~\eqref{eq: tt3} with zero and then solving with respect to $\phi_n$, we obtain the following critical point:
\begin{align}
c_n \triangleq \arctan\left(\frac{\Im\left\{\left[\widetilde{\v}_{\rm opt}^*\right]_n[\a]_n\right\}}{\Re\left\{\left[\widetilde{\v}_{\rm opt}^*\right]_n[\a]_n\right\}}\right).
\end{align}
The monotonicity of $f_1(\cdot)$ varies with the sign of $[\widetilde{\v}_{\rm opt}^*]_n$ and $[\a]_n$, which helps us to identify the phase shift $\phi_n$ as the critical or edge point that minimizes the norm difference \eqref{eq: f_W} relative to the $n$-th metamaterial, i.e.:
\begin{align}\label{eq: phi_1}
    \phi_n = \underset{\substack{x\in\{-\frac{\pi}{2},c_n,\frac{\pi}{2}\}}}{\arg\min}f_1(x).
\end{align}
To this end, we construct the diagonal matrix $\W_{\rm TA}$, and consequently $\W_{\rm RX}$, to optimally solve $\mathcal{OP}_1$ in the least-squares sense, since for the first-order approximation of \eqref{eq: BF}, the weights $w_n$ $\forall n$ in \eqref{eq: f_W} are decoupled.
\begin{figure*}
\begin{align}
    &f_1(\phi_n) \triangleq \left|[\widetilde{\v}_{\rm opt}]_n\right|^2 -\Im\left\{\left[\widetilde{\v}_{\rm opt}^*\right]_n[\a]_n\right\}-\cos(\phi_n)\Re\left\{\left[\widetilde{\v}_{\rm opt}^*\right]_n[\a]_n\right\}-\sin(\phi_n)\Im\left\{\left[\widetilde{\v}_{\rm opt}^*\right]_n[\a]_n\right\}+0.5|[\a]_n|^2\label{eq: tt2}\\
    &\qquad\qquad\qquad\qquad\frac{\partial f_1(\phi_n)}{\partial\phi_n}= \sin(\phi_n)\Re\left\{\left[\widetilde{\v}_{\rm opt}^*\right]_n[\a]_n\right\}-\cos(\phi_n)\Im\left\{\left[\widetilde{\v}_{\rm opt}^*\right]_n[\a]_n\right\}  \label{eq: tt3}
\end{align}
\vspace{-0.8cm}
\end{figure*}

\subsubsection{Second-Order Approximation}\label{eq: ex2}
The second-order approximation of $\W_{\rm RX}$ presented in Section~\ref{Sec: MC_approx} includes an addition term accounting for the mutual coupling, in particular, $\W_{\rm RX}\approx\P_{\rm SA}^{\rm H}(\W_{\rm TA}^{-1}-\W_{\rm TA}^{-1}\W_{\rm MC}\W_{\rm TA}^{-1})$. Analogous to Section~\ref{eq: ex1}, we set $\a = \P_{\rm SA}\v$ and reformulate the objective of $\mathcal{OP}_1$ in terms of the $n$-th diagonal value $w_n$, yielding:
\begin{align}\label{eq: f_2}
    \nonumber&f_2(w_n) \triangleq \left|\left[\widetilde{\v}_{\rm opt}-(\W_{\rm TA}^{-1}-\W_{\rm TA}^{-1}\W_{\rm MC}\W_{\rm TA}^{-1})^{\rm H}\P_{\rm SA}\v\right]_n\right|^2
    \\&=\left|\left[\widetilde{\v}_{\rm opt}\right]_n-w^*_n(1-\left[\W_{\rm MC}^{\rm H}\right]_{n,n}w^*_n)[\a]_n+w^*_nb\right|^2,
\end{align}
\begin{figure*}
\begin{align}
    x_1,x_2=\frac{[\a^*]_n-b^*\pm\sqrt{[\a^*]_n^2-2[\a^*]_nb^*-4\left[\widetilde{\v}_{\rm opt}^*\right]_n\left[\W_{\rm MC}\right]_{n,n}[\a^*]_n+(b^*)^2}}{2\left[\W_{\rm MC}\right]_{n,n}[\a^*]_n},\quad x_3=\frac{[\a^*]_n-b^*}{2\left[\W_{\rm MC}\right]_{n,n}[\a^*]_n}\label{eq: tt4}
\end{align}
\hrulefill
\end{figure*}
\hspace{-0.2cm}where $b\triangleq\sum_{j=1,j\neq n}^N\left[\W_{\rm MC}^{\rm H}\right]_{n,j}w_j^*[\a]_j$. Using a similar approach as before, we first compute $\frac{\partial f_2(\phi_n)}{\partial\phi_n}$ and then solve for $\phi_n$. For brevity, we omit the detailed calculations, which can readily be carried out using standard numerical tools. This yields the following critical points:
\begin{align}
    &c_{n,i} = \arctan\left({\frac{\Im\{x_i-0.5\}}{\Re\{x_i\}}}\right)\, \forall i=1,2,3
\end{align}
with $x_i$'s defined in~\eqref{eq: tt4}. We again choose the tunable phase shift $\phi_n$ as the critical or edge point that minimizes the norm difference \eqref{eq: f_2} with respect to the $n$-th metamaterial, i.e.:
\begin{align}\label{eq: phi_2}
    \phi_n =\hspace{-0.4cm}\underset{\substack{x\in\left\{-\frac{\pi}{2},\{c_{n,i}\}_{i=1}^{3},\frac{\pi}{2}\right\}}}{\arg\min}\hspace{-0.5cm}f_2(x).
\end{align}
Consequently, we construct the diagonal matrix $\W_{\rm TA}$ and, thus, $\W_{\rm RX}$ to minimize $\mathcal{OP}2$, while accounting for the mutual coupling effects between the DMA's metamaterials. However, optimality cannot be guaranteed, since the weights $w_n$ $\forall n$ are coupled in terms of $b$ in \eqref{eq: f_2}.  

The complete algorithm for computing the A/D BF weights of the DMA is summarized as follows: \textit{i)} solve for $\widetilde{\v}_{\rm opt}$ via the Rayleigh quotient maximization in $\mathcal{OP}$; \textit{ii)} Iteratively solve for $\v$ using the least squares closed form solution and, then, construct $\W_{\rm TA}^{-1}$ via \eqref{eq: phi_1} or \eqref{eq: phi_2} depending on the approximation order, until a stopping criterion is met. It is noted that closed-form solutions for the DMA's A/D BF weights can be derived for higher-order approximations of \eqref{eq: BF} analogously to Sections~\ref{eq: ex1} and~\ref{eq: ex2}. In those cases though, ensuring optimality is challenging, since the derivatives may produce multiple critical points.

\section{Numerical Results and Discussion}
In this section, we numerically evaluate the performance of the proposed RX DMA design for near-field localization, when used in conjunction with the maximum likelihood estimation scheme~\cite{myung2003tutorial}. We have simulated a system operating ats $28$ GHz with a $B=150$ KHz bandwidth, where the single-antenna UE was positioned in the vicinity of the RX DMA panel at $\theta = 30^{\circ}$, $\phi\in[0^{\circ}, 180^{\circ}]$, and $r\in[1,30]$ meters. In the DMA, the microstrips and metamaterials were spaced with $d_{\rm RF}=\lambda$ and $d_{\rm E}=\lambda/5$, respectively, while the rest of the RX DMA's modeling details were taken from \cite[Table II]{williams2022electromagnetic}. Each simulation was conducted over $500$ Monte Carlo simulations, with each including $T=200$ UE pilot transmissions, and the AWGN's variance was set as $\sigma^2=-174 + 10\log_{10}(B)$. Our localization approach has been compared with established partially-connected BF architectures. To facilitate comparisons we have optimized under the same localization objective,  the architectures  presented in~\cite{spawc2024} and~\cite{shu2018low} including an idealized DMA model, alongside a Hybrid A/D BF (HBF) structure. The inter-element spacing within each microstrip of the former was chosen as $\lambda/5$, whereas, for the latter, as $\lambda/2$. Notably, none of those model accounts for mutual coupling between adjacent antennas/metamaterials.
\begin{figure}[!t]
\centering
\includegraphics[scale=0.446]{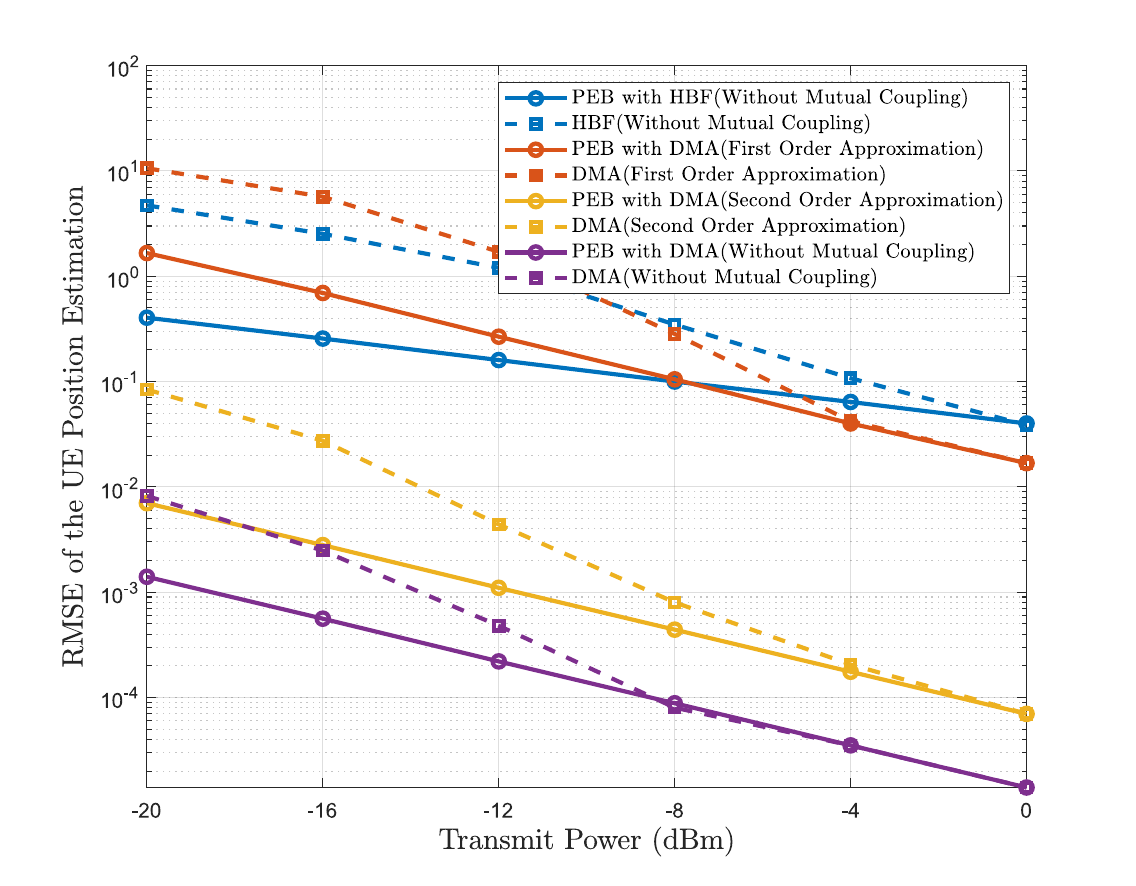}\vspace{-0.2cm}
\caption{\small{RMSE of the UE's position versus $P_{\rm max}$, considering an RX DMA equipped with $N_{\rm RF}=2$ microstrips, each hosting $N_{\rm E} = 256$ phase-tunable metamaterials. 
}}
\label{fig:RMSE}
\end{figure}

\begin{figure}[!t]
\centering
\includegraphics[scale=0.6]{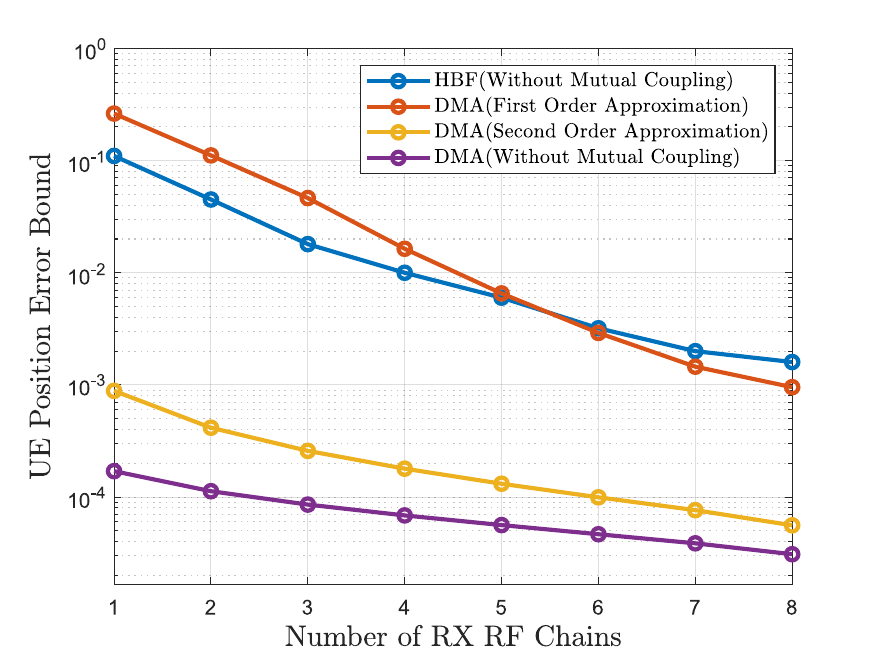}\vspace{-0.2cm}
\caption{\small{PEB of the UE's position versus the number of RX RF chains, considering an RX DMA with $N_{\rm E} = 256$ phase-tunable metamaterials per microstrip and $P_{\rm \max}=-12$ dBm.}}
\label{fig:PEB}
\end{figure}

In Fig.~\ref{fig:RMSE}, the Root Mean Square Error (RMSE) of the UE's position and the PEB performance with both presented approximated DMA models are illustrated for different partially-connected BF architecture with $N_{\rm RF} = 2$ RX RF chains, each attached to $N_{\rm E}=128$ antenna/metamaterial elements. As anticipated, all RMSE curves improve with increasing Signal-to-Noise Ratio (SNR) values, ultimately approaching toward their respective PEBs. It is shown that, despite accounting for mutual coupling, the proposed DMA-based models, demonstrate superior scalability with increasing SNR levels compared to traditional HBF architectures. This is attributed to the DMA denser element spacing, which ultimately allows them to surpass HBF's RMSE and PEB performance. Additionally, the second-order approximation model demonstrates adequate localization capability despite the burden of mutual coupling, but significant differences in the PEB and RMSE performance emerge between the idealized DMA model and the proposed ones. However, as it is shown in Fig.~\ref{fig:PEB}, these disparities primarily result from the limited number of RX RF chains employed.
In particular, the latter figure depicts the PEB performance of the considered BF architectures versus the number of RX RF chains in the given aperture, with each RX RF chain attached to $N_E = 256$ antenna/metamaterial elements. It can be observed that both DMA-based models demonstrate better scalability in terms of PEB performance compared to the HBF architecture, and outperform the HBF structure, despite the presence of mutual coupling. A shown, when the number of RX RF chains increases, the performance gap between the second-order approximated DMA model and the idealistic DMA model narrows. This trend highlights our solution's ability to deliver PEB performance that closely aligns with the ideal scenario. Furthermore, as the aperture's size increases, the first-order approximated model improves at a faster rate than the second-order one, making the former a strong candidate for XL antenna arrays, particularly due to its low complexity implementation. Thus, for such arrays, the proposed models achieve performance levels that are closer to the ideal case, contrasting with the results shown in Fig.~\ref{fig:RMSE}, where a significant performance gap was observed with a low number of RX RF chains for varying SNR values.

\section{Conclusion}
In this paper, we studied DMA-based reception design for UE localization, considering a state-of-the-art circuit-compliant model that accounts for mutual coupling within the analog BF stage. We presented two approximations for the analog BF matrix, whose accuracy was numerically assessed for various system parameters, to improve analytical tractability and enable efficient DMA optimization. A novel expression for the CRB for the estimation of the UE range and angular parameters as well as the PEB were derived, and the former was deployed as the minimization objective for the design of the RX DMA's A/D BF weights. We then capitalized on the DMA BF architecture to reformulate the localization objective as a Rayleigh quotient optimization problem, which allowed us to derive closed-form solutions for the A/D BF weights. Our simulation results verified the accuracy of our localization analysis, highlighting the advantages of our localization-optimized DMA design over well known XL antenna array architectures, which often neglect mutual coupling effects between their adjacent antenna elements. 

\section*{Acknowledgments}
This work has been supported by the SNS JU projects TERRAMETA and 6G-DISAC under the EU's Horizon Europe research and innovation programme under Grant Agreement numbers 101097101 and 101139130, respectively. 

\bibliographystyle{IEEEtran}
\bibliography{ms}
\end{document}